\documentclass[showpacs, two column]{revtex4}
\usepackage{graphicx}
\usepackage{amsmath}
\usepackage{appendix}

\begin{document}

\title{ Many-soliton bound states in dispersion-managed optical fiber: Possibility of fiber-optic transmission of three bits per clock period }

\author{Abdel\^{a}ali Boudjem\^{a}a}

\affiliation{
Department of Physics, Faculty of Sciences, Hassiba Benbouali University of Chlef, P.O.
Box 151, 02000, Ouled Fares, Chlef, Algeria.}

\begin{abstract}
We study the stability and the dynamics of many-soliton molecules in dispersion-managed (DM) optical fibers with focus on five-and seven-soliton molecules by analytical and numerical means.
In particular we calculate the binding force, pulse durations and equilibrium separations using a variational approach.
Predicted pulse shapes are in good agreement with those found by numerical simulations of the underlying nonlinear Schr\"odinger equation. 
Within limitations soliton molecules with up to seven solitons possibly allow to encode three bits of data per clock period.
\end{abstract}

\pacs{42.81.Dp, 42.65.Tg, 42.79.Sz}

\maketitle

\section{Introduction}

From time immemorial, information and communications have always formed the basis of human existence. 
Access to telecommunications such as telephones and the Internet is critical to the development of all aspects of a nation's economy.
Today, fiber-optic telecommunication systems play a major role in the advent of the Information Age mostly due to their very high data transmission capacity.
The dispersion phenomenon is a problem for high bit rate and long-haul optical communication systems. 
A solution of this problem might be soliton-molecules that have been suggested to increase the data-carrying capacity. 
Soliton based optical communication systems can be used over distances of several thousands of kilometers with huge information carrying capacity by using optical amplifiers. 

A few years ago, a stable bound state of two DM solitons in optical fibers called \textit{soliton molecule} was realized experimentally \cite {Mitschke2005} and very recently, the concept was extended by the
experimental demonstration of three-soliton molecules in dispersion-managed optical fibers by the same group \cite{Mitschke2012, Mitschke2013}. 
The phenomenon was also pursued theoretically in \cite {abdou, abdou1, Alm}.
The main motivation behind creating such molecules is to increase the bit-rate of data transfer in optical fibers beyond the binary regime with the advantage of the robustness of solitons in contrast to linear schemes. 
With four different symbols: logical zero, one (single soliton), two (two-soliton molecule) and three (three-soliton molecule), 
two bits of information can be coded in time slot (clock period).

The stability of soliton molecules attracted a great deal of interest \cite{Malom1, serhasg, Mart,Mart1, Nij, Gab, Shk, Mitschke2010, Turi}. 
The existence of a nonzero binding energy in terms of inter soliton separations in the molecule is an indication on its stability.
Most recently, the binding mechanism within a modified perturbation method is also discussed in \cite {Mitschke2013B} where it is found that a
multitude of equilibrium states, alternating stable and unstable can occur in special fiber configurations.
At the lowest possible separation a molecule is called to be in the ground state. 
Molecules at higher separations than the ground state are called higher-order soliton molecules \cite{Mitschke2013B}.

Motivated by these new prospects, we study the dynamics and the stability of many-soliton molecules 
by choosing appropriate parameters for the DM fiber.
Our aim in this paper is to check whether a fiber-optic transmission of three bits per clock period is possible and useful. 
Initially we develop a variational approximation to describe the dynamics and the binding mechanism of soliton molecules. 
The force of interaction between solitons (binding force) and equilibrium separations are calculated in the frame of this variational approach. 
The predictions of this method are compared with results of numerical simulations 
of the nonlinear Schr$\rm\ddot o$dinger equation (NLSE) and good agreement is found.  



\section{Dispersion-managed Nonlinear Schr{\"o}dinger Equation}  \label{introduction section}

Solitons in dissipative dispersion-managed optical fibers are described by the dispersion-managed nonlinear Schr$\rm{\ddot o}$dinger equation (DM NLSE):
\begin{equation}\label{gp1}
i\,E_z-{\beta_2(z)\over2}\,E_{tt}+S(z)|E|^2\,E=-i\,g(z)\,E ,
\end{equation}
where $E(t,z)$ $|E|^2$ [W], $z$ [m] and $t$[s] are the complex envelope of the electric field, the propagation distance, and the retarded time, respectively. 
$S(z)$ [1/(Wm)] and $g(z) $ [1/m] represent the nonlinearity and the gain/loss parameters, respectively. $\beta_2(z) [\text s^2/\text m]$ corresponds to the dispersion period defined
by
\begin{equation}
\beta_2(z)=\left\{\begin{array}{ll} \beta_2^+,\hspace{1cm}0\le z\le L^+,\\\beta_2^-,\hspace{1cm}L^+< z\le L^++L^-,
\end{array}\right.\label{doft}
\end{equation}
where $\beta_2^{+,-}$ are constant group velocity dispersion parameters of the respective fiber segments $L^{+,-}$. The length of the dispersion period is given by $L=L^+ + L^-$.

We use the transformation $E(t,z)=a(z)\,u(t,z)$, 
where $a(z)=a_0\exp{(-\int_0^z\, g(x)\,dx)}$ and  $a_0$ is dimensionless parameter.
This moves the loss term to the coefficient of the nonlinear term. Thus, Eq.(\ref{gp1}) reduces to 
\begin{equation}\label{gp2}
i\,u_z-{\beta_2(z)\over2}\,u_{tt}+\gamma (z)|u|^2\,u=0 ,
\end{equation}
where $\gamma (z)=S(z) a(z)^2$ is the fiber's effective nonlinearity.

In general, for any $\gamma(z)>0$, we then introduce a new coordinate $z'[1/W]$ defined by $z'(z)=\exp{(-\int_0^z\, \gamma(x)\,dx)}$ 
\begin{equation}\label{gp3}
i\,u_{z'}-{\beta_2'(z')\over2}\,u_{tt}+|u|^2\,u=0 ,
\end{equation}
where $\beta_2'(z')=\beta_2(z')/\gamma(z') [\text W. \,\text s^2]$ represents fiber’s effective dispersion including the variations both of fiber’s GVD
and effective nonlinearity.\\
Equation (\ref {gp3}) permits us to describe the variations of dispersion, nonlinearity, and optical power due
to loss or gain by a single variable on the distance which is measured with the accumulation of nonlinearity. For a negative
constant, (\ref {gp3}) is called the NLSE and can be analytically solved for any initial input by using the inverse scattering transformation having soliton solutions \cite {Zaka, akira}.

For numerical purposes, it is useful to set Eq.~(\ref{gp3}) into a dimensionless form. 
First we introduce $Z=z'/L'$, $T=t/\tau_m$ and $Q (Z,T)=u(z',t).\,\sqrt{L'}$ where $\tau_m$ is
the characteristic time scale equal to the pulse duration of the laser source and $L'=L^{+}\gamma^{+}+L^{-}\gamma^-$ is the length
of the dispersion map period. 
In terms of these parameters, the dimensionless NLSE takes the form
 \begin{equation}\label{dd}
iQ_Z -\frac{D(Z)}{2}Q_{TT} +|Q|^2 Q=0,
\end{equation}
where $D=\beta_2^\prime L'/ \tau^2_{m}$.\\ 

\section{Variational approach}\label{var sec}

The variational approach is proved to be efficient for the analysis of non-integrable soliton bearing equations in different areas of physics \cite {Tur}.\\
We then employ the following trial wavefunction 
 \begin{equation}\label{trial}
Q(Z,T)=A\sum\limits_{j=1}^ {N} \exp\left[-\frac{(T-\eta_j)^2}{q^2}+i \alpha_j (T-\eta_j)^2+i\varphi_j\right],
\end{equation}
where $A$  guarantees the normalization of $Q$ to the number of solitons in the molecule, namely $N=7$.
The variational parameters $q(Z)$, $\varphi (Z)$, $\eta (Z)$ and $\alpha(Z)$ correspond respectively to the width, the relative phase, 
the peak position and the chirp of the soliton. 


The Lagrangian corresponding to Eq.(\ref{dd}) reads
\begin{equation}\label{Langr}
L [Q,Q^*]= \int_{-\infty}^\infty  \frac{i}{2} \left(QQ_Z^*-Q^*Q_Z\right) dT-E,
\end{equation}
where 
 \begin{equation}\label{enr}
E=-\int_{-\infty}^\infty \left[\frac{D}{2} |Q_T|^2+\frac{1}{2} |Q|^4 \right]dT,
\end{equation}
is the energy functional. 

For the sake of completeness we start with the development of the variational approach for a single DM soliton. The energy of pulse is given by
\begin{equation}\label{En}
E_s=\int_{-\infty}^\infty |Q|^2 dT=A^2q \sqrt{\pi/2}.
\end{equation}
In  real units, this  energy can be expressed  as $E[J]= (\tau_m/L') {\cal E}$ with ${\cal E}$ [J] is the original molecule's energy.
Equations of motion for the differents variational parameters are then derived from the Euler equations, namely
\begin{equation}\label{EM}
\frac{d}{dZ}.\frac{\partial L}{\partial \psi_Z}-\frac{\partial L}{\partial \psi}=0,
\end{equation}
where $\psi$ and $\psi_Z$ denote a variational parameter and its spatial derivative, respectively.
Taking  the corresponding derivatives, the Lagrangian (\ref{Langr}) of single soliton reads
\begin{align}\label{L1}
L=E_s \left[\frac{q^2}{4}\alpha_Z+\frac{1}{2} \varphi_Z-\frac{D(Z)}{2} \left (\frac{1}{q^2}+\alpha^2q^2\right)-\frac{E_s}{2 \sqrt{\pi} q}\right].
\end{align}
Using (\ref{EM}), equations of motion for variational parameters take the following explicit form:
\begin{subequations}\label{E:EM}
\begin{align} 
& E_Z=0,  \label{E:EM1} \\  
&\alpha_Z+ 2D(Z) \left(\frac{1}{q^4}-\alpha^2\right)+\frac{E_s}{\sqrt{\pi} q^3}=0,  \label{E:EM2} \\ 
&q_Z+2 D(Z) \alpha q=0.  \label{E:EM3}
\end{align}
\end{subequations}
Interestingly, Eq.(\ref{E:EM1}) shows the conservation of the energy while Eqs. (\ref {E:EM2}) and (\ref {E:EM3}) determine the width $q$ and the chirp $\alpha$ of a single DM soliton. 
Combining the last two equations, we obtain the following equation for the pulse width
\begin{align} \label{Pot}
q_{ZZ}=F(q)=-\frac{\partial U}{\partial q}, 
\end{align}
where
$$ U=  \frac{2D(Z)^2}{q^2}-\frac{2D(Z) E_s}{\sqrt{\pi} q}.$$
The shape of $U(q)$ gives insights on the robustness of solitons and the binding mechanism between in the molecules.\\
For the two- and higher order- soliton molecules, the resulting coupled equations of motion, which turn out to be lengthy and hence will not
be shown here for convenience, will then be solved numerically. \\
We define the relative soliton separation $\Delta=\eta_j-\eta_i$, the relative centre-of-mass position $\xi=(\eta_j+\eta_i)/2$ and the relative phase $\phi=\varphi_j-\varphi_i$.
Note that different relative phases lead to an asymmetric energy transfer between adjacent solitons which may eventually cause the molecule to disintegrate \cite{Mitschke2013B}. 
The DM fiber parameters are given in Tab.~\ref{tab:solmoltab}.
\begin{table}[h]
    \begin{displaymath}
    \begin{array}[t]{c}
\hline
        \begin{array}[t]{rcl @{\quad  \quad} rcl}
            \beta_2^-       &=& -4.7~{\rm ps}^2 \, {\rm km}^{-1}     & \beta_2^+     &=& 3.9~{\rm ps}^2 \, {\rm km}^{-1}\\
            \gamma^-        &=& 1.9~{\rm W}^{-1} \, {\rm km^{-1}} & \gamma^+      &=& 1.9~{\rm W^{-1}km^{-1}}     \\
            L^-             &=& 26.5~{\rm m}                        & L^-           &=& 24.5~{\rm m}                  \\
        \end{array}
    \\ \hline
        \begin{array}{rcl}
            \overline{\beta}_2          &=& -0.57~{\rm ps}^2 \, {\rm km}^{-1}        \\
            \overline{\gamma}           &=& 1.9~{\rm W^{-1} \, {\rm km}^{-1}}       \\
            L                           &=& 51.0~{\rm m}                  \\
        \end{array}                                  \\
\hline
    \end{array}
    \end{displaymath}
    \caption{Parameters used for the simulation of ground state soliton molecules}
    \label{tab:solmoltab}
\end{table}

\begin{figure} [htb]						
  \centering
\includegraphics[scale=0.8] {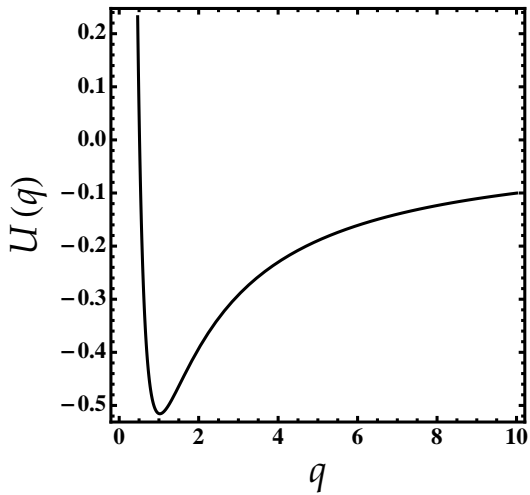}
  \caption{ The potential $U$ from Eq.(\ref{Pot}) as a function of the pulse width $q$ for $N=7$,  $\eta_1=-6,\eta_2=-4, \eta_3=-2, \eta_4=0, \eta_5=2, \eta_6=4, \eta_7=6$,
the phase alternates as 0, $\pi$, and chirp parameter $\alpha =0.01$. The rest of the parameters are indicated in Tab.\ref{tab:solmoltab}.}
  \label{bind1}
\end{figure}

Figure~\ref{bind1} shows that the potential of seven equidistant DM solitons has a local minimum at  $q \sim 1 $.
For $q \lesssim 1 $,  the initial force is repulsive, and thus, the solitons may diverge from each other leading to the dissociation of the molecule.
Increasing the pulse energy, the strength of the bond in the molecule grows indicating that the DM soliton-molecule becomes more robust.




\begin{figure}	 [hbt]		
 \centering
 \includegraphics[scale=0.4]{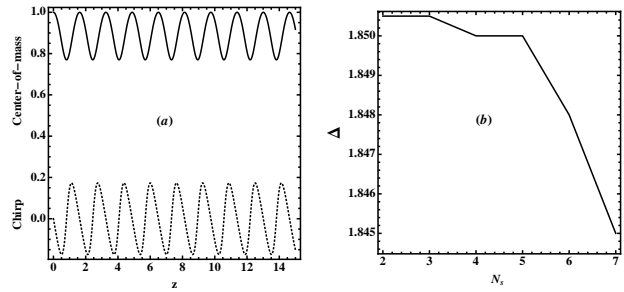}
  \caption{\label{mcp1}  (a) Evolution of the center-of-mass positions (top) and the chirp parameter (bottom).
(b) The inter-soliton separation as a function of the number of solitons for $q=1$. Parameters are the same as in Fig.\ref{bind1}.}
\label {Separ}
\end{figure}

Figure.\ref{Separ}.a  depicts that the center of mass and the chirp exhibit periodic dynamics which may reduce/increase the temporal separation between the center-of-mass positions
of solitons. 
This is in fact, qualitatively confirmed in Fig.\ref{Separ}.b where we clearly observe that until the third soliton, interpulse separation remains stable yielding a robust molecule.
Whereas from fourth to seventh soliton the separation is slightly lowered and increased due to the inherent repulsive force between solitons.
In such a situation, the pulse performes oscillations with frequency  that can be evaluated variationally by expanding the potential  (\ref{Pot}) near its
equilibrium value $\omega_0= \sqrt{-\partial F/\partial q |_{q=q_0}} \simeq 12.6$.

One can conclude that the soliton separation in a molecule does not has not a unique value but instead changes with the number of solitons $N$ \cite {abdou} 
and other factors such as the chirp. 
In some cases higher-order equilibrium separations appear which is in good agreement with numerical results of \cite{Mitschke2013B}.

\section{Numerical results}\label{Num sec}

To verify our variational results, we solve numerically the NLSE (\ref{dd}) for the five-soliton molecule and the seven-soliton molecule.
Numerical simulations were performed using the split-step Fourier transform method, with $2048$ Fourier modes and a normalized step size of $dz = 5 \cdot 10^{-4}$.

\begin{figure} [htb]						
  \centering
\includegraphics [scale=0.4,clip] {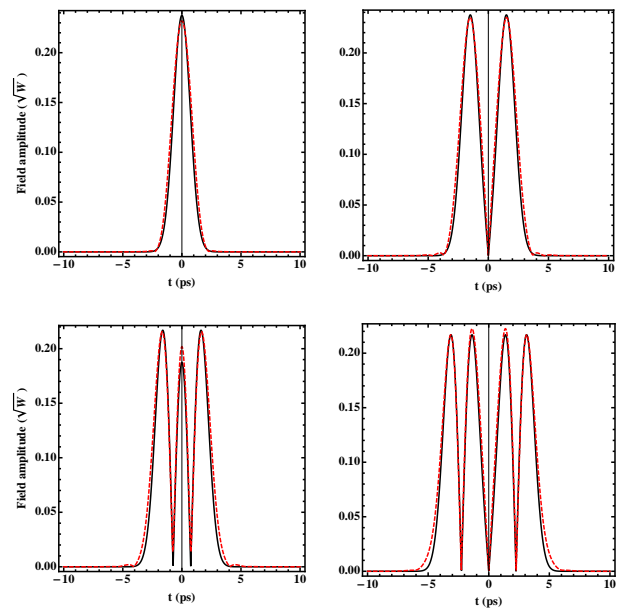}
  \caption{ Field amplitude envelopes along dispersion-managed fiber of 1-soliton (top left), 2-soliton (top right), 3-soliton (bottom left) and 4-soliton molecule (bottom right).
 Solid line: variational calculation and red dashed line: numerical simulation.
Parameters are the same as in Fig.\ref{bind1}.}
  \label{Profile}
\end{figure}

In Fig. \ref{Profile} we compare the intensity profile obtained by our
variational calculation with the  simulation results for single soliton and 2-,3-, and 4-soliton molecules.
The figure shows a good agreement between the variational and the direct numerical solution of the NLSE.
The intensity profile was calculated using the equilibrium conditions found by minimizing the energy functional.

\begin{figure} [htb]				
  \centering
  \includegraphics [scale=0.5,clip] {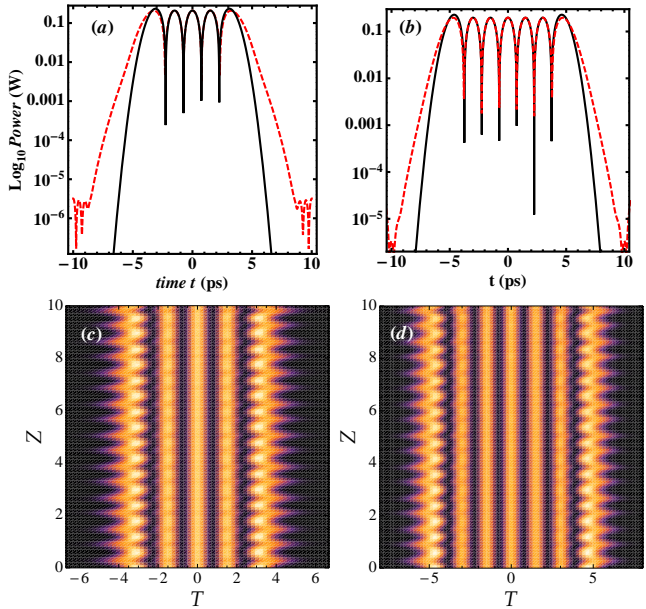}
  \caption{(Color online) Field amplitude envelopes along dispersion-managed fiber
  of a five-soliton molecule (panel a), and seven-soliton molecule (panel b) in logarithmic scale for $\Delta= 1.85$, $q=1$ and $\alpha=0$.
  Solid line: our variational calculation and red dashed line: results of Nijhof's method \cite{Nij} applied to the original NLSE. 
	Lower panels: Corresponding propagation of the five-soliton molecule (panel c) and seven-soliton molecule (panel d) over 8 dispersion periods.}
  \label{Many}
\end{figure}

Figs.~\ref{Many} (a) and (b) show the power profiles of five- and seven-soliton molecules on logarithmic scale. 
Results from our theoretical treatment (solid black curve) are compared to numerical solutions (red dashed curve) obtained from Nijhof's method \cite{Nij} 
applied to the original NLSE (\ref {gp3}).
Slight deviations of the two waveforms are noticeable only at the outer solitons and in far tails.
There are slightly different peak powers of the individual solitons; outer solitons are higher than the inner ones in agreement with our recent theoretical predictions \cite {abdou}. 

In Figs.~\ref{Many} (c) and (d) the propagation of the considered long soliton molecules is shown. 
The narrow separation results in a strong inter soliton interaction but solitons can travel together in the same cell and maintain their respective positions.
These ground-state soliton molecules survive here only over 8 dispersion periods or less and disintegrate very fast for longer soliton trains.
Different peak powers result in different nonlinear phase offset evolutions. 
This will corrupt the relative phase relation between the solitons in the molecule and can lead to an energy transfer between the solitons resulting in a collapse of the structure.
This is in agreement with numerical results of Hause et al. \cite{Mitschke2013B} where it was shown that three-soliton molecules at the ground state collapse after a few tens of dispersion periods
while higher-order three-soliton molecules can survive on long-haul distances. 
Indeed, solitons of higher-order soliton molecules interact only with the next-neighbor and the higher separation results in a weaker interaction force due to the smaller pulse overlap. 
Therefore this type of soliton molecules can travel on long-haul distances and hence, it can arranged in long chains at higher-order separations.


Single solitons can not be arranged at any narrow separation to form a long soliton molecule. 
For example Fig.~\ref{mcp1} depicts that when seven Gaussian pulses are placed at very narrow temporal separations an initial double pulse structure due to the interference is formed.
Thus, here an approximative in-phase two-soliton molecule is formed.

\begin{figure}[hbt]			
 \centering
 \includegraphics [scale=.7,clip] {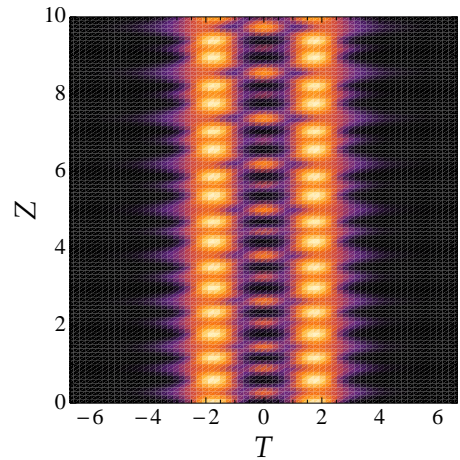}
  \caption{\label{mcp1}  Propagation of seven solitons very close to each other ($\Delta =0.5 $) results in a double pulse structure.
	The parameters are the same as in Fig. \ref{Many}.}
\end{figure}

\section{Conclusion} \label{conc}

In conclusion, in  this work we have considered many-soliton molecules propagating in optical dispersion-managed fibers. 
Using a variational treatment, the binding force, equilibrium separation and the chirp of soliton molecules was calculated with emphasis on five-soliton molecules and seven-soliton molecules. 
We have solved numerically the NLSE for comparison and it is found that the field amplitudes of the obtained trial functions agreed favorably with the numerical molecule solution.
It is further confirmed that the equilibrium separation between dispersion-managed solitons does not have a unique value.
There is a ground state separation at the narrowest separation and higher-order states at larger separations in good agreement with semianalytical results of \cite{Mitschke2013B}.
The experimental achievement of many-soliton molecules remains challenging and would open fascinating prospects
for fiber-optic transmission of more than two bits per clock period.


\section{Acknowledgements}
We are grateful to Alexander Hause, Fedor Mitschke and Usama Al-Khawaja for stimulating discussions and for useful comments on the manuscript.
We acknowledge support of the Algerian government under the research Grants No. CNEPRU-D00720130045. 

\end{document}